\documentclass[%
  preprint,
superscriptaddress,
 amsmath,amssymb,
 aps,
prl,
]{revtex4-2}

\usepackage{graphicx}
\usepackage{dcolumn}
\usepackage{bm}
\usepackage{xcolor}
\usepackage{multirow}

\begin{document}

\title{Machine Learning Potential of a Single Pendulum}

\author{Swarnendu Mandal}
\email{swarnenduphy35@gmail.com}
\affiliation{Central University of Rajasthan, Ajmer, Rajasthan, India - 305817}

\author{Sudeshna Sinha}
\email{sudeshna@iisermohali.ac.in}
\affiliation{Indian Institute of Science Education and Research Mohali, Punjab, India - 140306}

\author{Manish Dev Shrimali}
\email{shrimali@curaj.ac.in}
\affiliation{Central University of Rajasthan, Ajmer, Rajasthan, India - 305817}

\begin{abstract}
Reservoir Computing offers a great computational framework where a physical system can directly be used as computational substrate. Typically a ``reservoir'' is comprised of a large number of dynamical systems, and is consequently high-dimensional. In this work, we use just a {\em single} simple low-dimensional dynamical system, namely a driven pendulum, as a potential reservoir to implement reservoir computing.  Remarkably we demonstrate, through numerical simulations, as well as a proof-of-principle experimental realization, that one can successfully perform learning tasks using this single system. The underlying idea is to utilize the rich intrinsic dynamical patterns of the driven pendulum, especially the transient dynamics which has so far been an untapped resource. This allows even a single system to serve as a suitable candidate for a ``reservoir''. Specifically, we analyze the performance of the single pendulum reservoir for two classes of tasks: temporal and non-temporal data processing. The accuracy and robustness of the performance exhibited by this minimal one-node reservoir in implementing these tasks strongly suggest a new direction in designing the reservoir layer from the point of view of efficient applications. Further, the simplicity of our learning system offers an opportunity to better understand the framework of reservoir computing in general and indicates the remarkable machine learning potential of even a single simple nonlinear system.
\end{abstract}

\maketitle


The ability of dynamical systems to process information has commanded long-standing interdisciplinary research interest\cite{shaw1981strange,sinha1998dynamics,ieee}. 
There are several examples of natural systems with the capability to perform different forms of \textit{intrinsic computation}\cite{crutchfield2010introduction,mainzer2007thinking,toffoli2004nothing}. 
In the context of machine learning, the overarching question is how ideas from physics or physical systems can enhance existing concepts. On one hand, research directions that can enhance the performance of algorithms in handling data from dynamical systems is a very pertinent question \cite{pre2020,csfx2020,nonlineardynamics2021,lai,trichygroup1}. On the other hand, research efforts to utilize physical systems to implement machine-learning learning algorithms have serious implications for new concepts in the field of artificial intelligence. This line of enquiry also has consequences for gauging the information processing capacity of naturally-occurring or human-engineered physical, chemical, and biological systems \cite{Beniaguev613141}.

Here we consider the Reservoir Computing (RC) technique to exploit a dynamical system for machine learning. RC is a recurrent neural network (RNN) based computational framework, in which the memory capability and rich dynamics of an RNN can be used for computation without actually training the network structure itself. Instead, training the readout layer is sufficient to achieve good performance \cite{jaeger2001echo,maass2002real}. In this framework, the network is called the \textit{reservoir}, as it stores the input as a high dimensional Spatio-temporal pattern, such that a linear transformation can efficiently extract the desired output in \textit{readout}. Formally, a low-dimensional temporal input $u(t)$ is transformed into a much higher dimensional state vector $x(t)$ by the reservoir. These state vectors are processed further by the linear readout to get a desired output. For its simplicity, scalability and lower training costs, reservoir computing has attracted widespread research interest, both in terms of applications \cite{lukovsevivcius2009reservoir,lukovsevivcius2012reservoir,tanaka2019recent,nakajima2020physical,pathak2018model,zhong2021dynamic,rafayelyan2020large,ghosh2021reservoir,saha2020predicting,zhang2021learning} as well as basic development of the general framework \cite{vlachas2020backpropagation,carroll2019network,silva2021reservoir,carroll2020reservoir}. 



\par 
In this letter, we show through both numerical simulations and experimentation, that a surprisingly simple system, namely a single forced pendulum has sufficient richness in its dynamics to process information for intelligent computation. The central idea is that, instead of multiplexing the input in state space, we encode the inputs in the temporal patterns, effectively making it act like a high dimensional system \cite{jensen2017reservoir,appeltant2011information}. 
In this work, we will assess the performance of our reservoir in 
the arena of both temporal and non-temporal tasks, and demonstrate that both these classes of tasks can be performed using our minimal one-node reservoir with strikingly good performance.

\medskip

\begin{figure}
	\centering
	\includegraphics[scale=0.45]{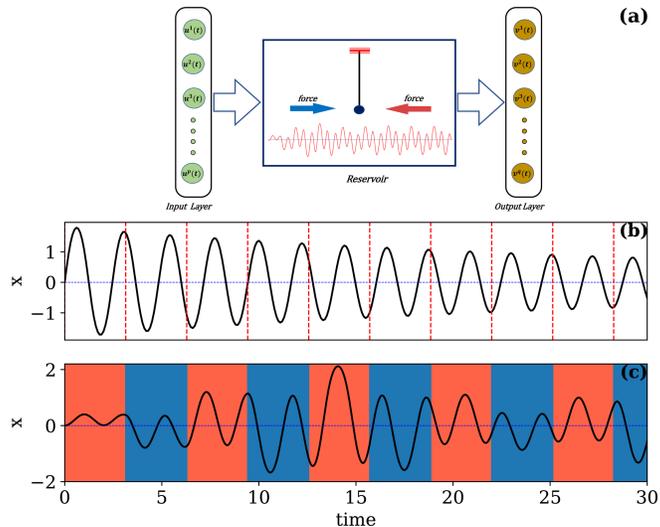}
	\caption{\textbf{(a)} Schematic of the reservoir; \textbf{(b)} dynamics of an under-damped pendulum without forcing (solid black line) with the 
	period of the driving force shown by dotted red lines; \textbf{(c)} Transient dynamics of the system (solid black line) 
	under periodic forcing (indicated by alternating red and blue backgrounds). In the presence of periodic driving, the dynamics is quasiperiodic here, as the frequency of force is not equal to the natural frequency of the pendulum. }
	\label{Fig_scheme}
	\end{figure}

\noindent
{\bf Reservoir Dynamics:} \ Specifically, we consider a pendulum of length $l$ with a bob of mass $m$, periodically driven with a force of amplitude $F$, depicted schematically in Fig.~\ref{Fig_scheme}. Given a  damping coefficient $b$ of the medium, the equation of motion can be written as

	\begin{align}
	\frac{d^2x}{dt^2} = -\frac{g}{l}sin(x) - k\frac{dx}{dt}+ f~sign~[sin(\omega t)].
	\end{align}
	where the $sign[\cdot]$ function represents a square wave that toggles between $+1$ and $-1$ when the argument switches from positive to negative.
	Here $f = \frac{F}{m}$ and $k = \frac{b}{m}$ are the amplitude of the force and damping coefficient per unit mass. This system can yield quasiperiodicity (as depicted in Fig.~\ref{Fig_scheme}) and has been studied extensively \cite{PhysRevA.35.4404,PhysRevLett.55.2103,PhysRevA.39.2593}. With no loss of generality, we have considered $k = 5\times 10^{-2}$ and $l=1.0$ for the numerical simulations presented in this study.

\par
\begin{figure}
	\centering
	\includegraphics[scale=0.4]{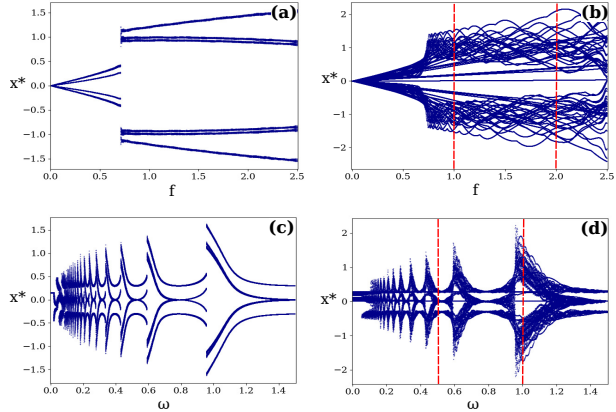}
	\caption{Bifurcation Diagram of the reservoir dynamics,  \textbf{(a-b)} with respect to amplitude $f$ of the driving force, with $\omega = 1.0$, and \textbf{(c-d)} with respect to the driving frequency $\omega$, with $f=1.5$. The first column \textbf{(a), (c)} represents the asymptotic dynamics, and the second column \textbf{(b), (d)} shows the transient dynamics starting from initial state $[x,\dot{x}]=[0,0]$. The region between dashed red lines are used to encode the input to the reservoir.}
	\label{Fig_bif_pend}
	\end{figure}

 The unique dynamics arising at each point in $f-\omega$ parameter space is evident from the bifurcation diagrams shown in Fig.~\ref{Fig_bif_pend}.
Unlike most studies, here we examine the temporal patterns arising, not just in the asymptotic case, but in the transient dynamics as well.
The comparative features of the transient reservoir dynamics and the asymptotic dynamics can be seen from the two columns displayed in the figure. Clearly, the {\em transient dynamics provides a richer repertoire of nonlinear patterns than the asymptotic behaviour, and we will crucially use this aspect to encode information more efficiently.}

\medskip
	
\noindent	
{\bf Input Encoding}: \ Multiplexing the input signal efficiently into the reservoir dynamics is a crucial step for reservoir computing. The complete information should be stored into the reservoir. For our case, we have three possible choices to feed the input to the system. One option is to encode the input information with the initial condition. But this is not an efficient choice. As different trajectories can evolve to the same attractor, the pre-images are not unique after transience, with different initial conditions  producing the same asymptotic trajectories. Hence this will lead to input information loss and hinder robust and consistent input encoding. Alternately, one can encode the input using the two system parameters: amplitude ($f$), and frequency ($\omega$) of the applied force. Either of these two options is a better choice than input encoding with initial conditions, as each point in $f-\omega$ space gives rise to an unique dynamical sequence.

In this work we focus on tasks involving one-dimensional inputs, and hence only one parameter needs to be varied for input encoding, while the other parameter can be kept fixed. We will consider input-encoding using both the amplitude and the frequency of  forcing,
in order to compare the efficacy of these two alternate methods of input-encoding for different tasks, including their robustness in the presence of noise. 
We will denote the scheme of input-encoding using forcing amplitude $f$ as \textit{amplitude encoding}, while the scheme where inputs are encoded  using the forcing frequency $\omega$ will be simply referred to as \textit{frequency encoding}.

First consider the \textit{amplitude encoding} scheme where we multiplex the input with the amplitude of force $f$. In this scheme we need to choose a range of the parameter, say $f \in [f_{min}, f_{max}]$, and we then need to scale all input points into this range. This scaling transformation ($u \rightarrow f$) can be expressed as $f = f_{min} + (f_{max}-f_{min})u$, where, $u$ is the normalized input in range $[0,1]$. Formally, $u=\frac{min[\tilde{u}(t)]+\tilde{u}(t)}{max[\tilde{u}(t)]-min[\tilde{u}(t)]}$, $\tilde{u}(t)$ being the original input signal. Specifically, the range of $f$ for this scheme is taken to be [1,2] as shown in Fig.~\ref{Fig_bif_pend}(b). For the  \textit{frequency encoding} scheme, we can proceed in a similar fashion, with parameter $\omega$ replacing parameter $f$ in the formalism.
	
\par As a test-bed to gauge the performance of our system we will consider two distinct classes of tasks: one task will involve processing non-temporal signals and another task will consider processing temporal signals.
In general, Reservoir Computing has proven to be successful in solving time dependent data processing, stemming from the nonlinear memory effect of the reservoir. But for non-temporal tasks, we need to remove the memory effect. This can be achieved by resetting the reservoir to a fixed point after feeding an input data point to it. Specifically for our case, the reservoir is set to $[x,\dot{x}]=[0,0]$ 
after each input.
	
\medskip
	
\noindent	
{\bf Reservoir State and Regression:} \ The transient dynamics of the reservoir is stored in a discretized form as the state vector. Only the transient part of the dynamics is considered as it produces richer nonlinear repertoire than its asymptotic behavior. More formally, the state vector is the set of variables $x(t)$ recorded at a fixed sampling rate $\kappa\Omega~[\kappa=1,2,3...]$, an integer multiple of the sampling cycle frequency $\Omega$, i.e. for each sampling cycle we record $\kappa$ values of $x(t)$. Thus if we store the data for $N$ cycles for each input, states to be stored can be written as $S = [x(0),~x(\tau),~x(2\tau),~....,~x(\kappa N\tau)]$, where $\tau = \frac{2\pi}{\kappa\Omega}$ is the sampling interval. For each point $\tilde{u}(t_i),~i=1,2,3,...$ of input signal, one state vector $X_i$ is formed from $S$.

Now, forming the reservoir state vector $X_i$ from stored states $S$ is different for temporal and non-temporal tasks. For non-temporal tasks, 
we produce the state vector as the column matrix $X_i = [S_i]^T$, where $[\cdot]^T$ represents the transpose. 
For temporal tasks, we form the state vector corresponding to any particular input with the current states as well as states corresponding to certain number of previous states, i.e. we take \ $X_i = [w_0S_{i-m},~w_1S_{i-m-1},~......,~w_{m-1}S_{i-1},~w_mS_i]^T$, where $w_j,~j=0,1,2...,m$ are the weights of previous input states, following a linear distribution in the range $[0,1]$. Here $m$ is the finite \textit{memory}, which can be considered as a hyper-parameter to be tuned for different kinds of temporal tasks, allows us to achieve the required \textit{fading memory} to process temporal data. In our numerical simulations, we have considered $m=100$.

For the two schemes of input encoding the value of $\Omega$ is different. In the \textit{amplitude encoding} scheme we take $\Omega = \omega$, the frequency of driving force and for the \textit{frequency encoding} scheme we consider $\Omega = \omega_0$, the natural frequency of the oscillator. In general, $\Omega$ can be treated as another hyper-parameter for both schemes.

 Thus, for a complete input signal $\tilde{u}(t)$ one has the reservoir state vector matrix $\Re = [X_1,~X_2,~X_3,...,X_L]$, $L$ being the length of input signal. So the matrix $\Re$ has the dimension $\kappa N\times L$ for Task - I and $m\kappa N\times L$ for Task-II, where $\kappa$ and $N$ are hyper-parameters that can be optimized for best results.

\par Now if the corresponding output for $\tilde{u}(t)$ is $\tilde{v}(t)$, the linear transformation between the output signal and the reservoir state vector matrix can be written as $\tilde{v} = W\Re,$ where, $W$ is the $1\times \kappa N$ dimensional connection matrix. This matrix can be evaluated using training data set by regression method as $W = \tilde{v}\Re^{-1}$. Specifically for this purpose, we have used the \textit{Moore-Penrose pseudoinverse} \cite{barata2012moore}.

\medskip

\noindent
{\bf Machine Learning Tasks:} \ To check the performance of the reservoir we consider two tasks. The first task is non-temporal, and involves the learning of a high-degree polynomial. The second task involves temporal data processing, and considers the difficult task of using data from one state variable to infer another state variable in a chaotic system.

Specifically, the aim of Task - I is to approximate a $7^th$ degree polynomial given by $f(x) = (x-3)(x-2)(x-1)x(x+1)(x+2)(x+3)$ in the range $x\in[-3,3]$. As this task corresponds to non-temporal input processing, one input point $x$ is necessary and sufficient to get the corresponding output $f(x)$.

Our second task (Task - II) pertains to the  reconstruction of a chaotic attractor dealing with temporal data processing. As an illustrative example, here we consider the state variable $x(t)$ of a chaotic Lorenz system $[\dot{x} = 10(y-x),~\dot{y} = x(28-z)-y,~\dot{z} = xy - 8z/3]$ as input to infer another state variable $y(t)$ in output
\cite{lorenz1963deterministic,sparrow2012lorenz}.


\medskip

\begin{figure}[ht]
	\includegraphics[scale=0.4]{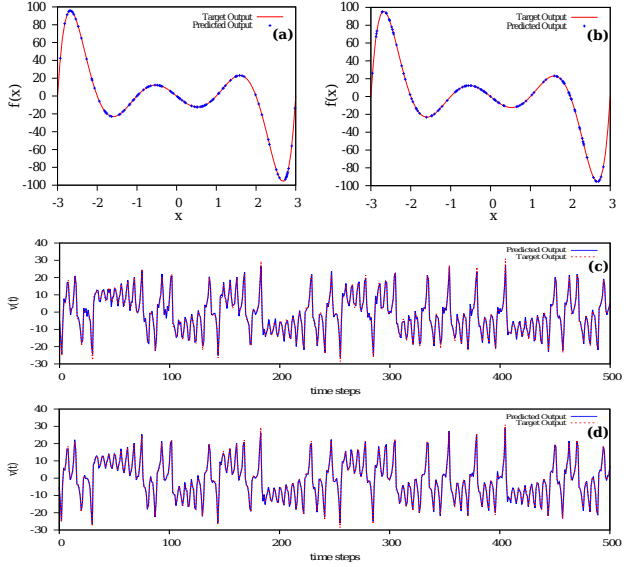}
	\caption{\textbf{(a)} The comparison of predicted output with target for Task-I \textbf{(a - b)} and Task-II \textbf{(c - d)}. \textbf{(a), (c)}  are the results obtained with the {\em amplitude encoding} scheme and \textbf{(b), (d)} are those obtained with {\em frequency encoding}.}
	\label{Fig_result}
	\end{figure}

\noindent
{\bf Results:} \ The efficiency of the reservoir is analysed by estimating the accuracy of the tasks it performs, quantified by root mean square error (RMSE) of the predicted output with reference to the target one. We find that the reservoir works with great accuracy for both temporal and non-temporal tasks, for both the schemes, as is clearly discernible from Fig.~\ref{Fig_result}. The success of our single-node reservoir is also evident quantitatively from Table \ref{Tab_noise}, which lists 
the order of accuracy obtained for the tasks.

For Task-I, we find that a reservoir trained with only $500$ data points, can approximate the polynomial with RMSE of the order of $10^{-10}$. Further, smaller training data sets does not significantly degrade the accuracy obtained. For instance, even training data sets with 
size as low as $100$, yields accuracy of the order of $10^{-6}$.

\par For the case of Task-II, the reservoir was trained with data set of length $5000$, and it yielded an accuracy of the order of $10^{-3}$. For the tasks we have considered, the prediction accuracy of the trained reservoir is independent of the testing data length, and for the numerical results listed in Table~\ref{Tab_noise} we have taken the testing data length to be the same as the training data length. So from the results it is clear that even for the tasks involving intensive and complex information processing, the one-node reservoir predicts the output successfully.

\bigskip

\noindent
{\bf Performance in Presence of Noise:} \ We now assess the robustness of the performance in the presence of a noise floor. In order to examine the effect of noise on the performance, we have perturbed each state variable with a random noise, uniformly distributed in the range $[-0.01:0.01$]. The results are displayed in Table \ref{Tab_noise}. It is clear that the performance is reasonably stable even in the presence of such significantly large noise. Further, we notice that encoding inputs via the frequency of the drive is more robust and accurate than encoding inputs via the amplitude of forcing. This suggests that for optimal and most robust implementation, different control parameters for encoding information should be investigated, as the nature of the dynamics could be quite different under variation of different parameters, leading to different robustness in the presence of noise.

\begin{table}
\centering
\begin{tabular}{|c|c|c|}
\hline
Input encoding & Task - I & Task - II \\ 
\hline
$f$  & $ 10^{-10}$/\color{red}{$10^{-2}$} & $10^{-3}$/\color{red}{$10^{-3}$} \\ \hline
$\omega$ & $ 10^{-8}$/\color{red}{$10^{-3}$} & $10^{-3}$/\color{red}{$10^{-3}$} \\ \hline
\end{tabular}
    \caption{Comparison of performance, as quantified by RMSE, under two different schemes of input encoding, with one method using the forcing amplitude $f$ and the other method using forcing frequency $\omega$ to encode inputs. 
The first value reports the order of the RMSE obtained from noise-free systems, while the second value (marked in red) gives the results obtained in the presence of noise.
}
    \label{Tab_noise}
\end{table}

\bigskip

\noindent
{\bf Proof-of-principle experiment:} \ We have also investigated the performance of a single-node reservoir that utilizes actual laboratory data from an experimental realization of a forced pendulum. Remarkably, even this simple experimental system yields very good performance, as seen from the results displayed in Fig.~\ref{Fig_expt}. The detailed discussion of experimental setup and procedure is listed in appendix \textit{B} of the supplementary material.


\begin{figure}
    \centering
    \includegraphics[scale=0.6]{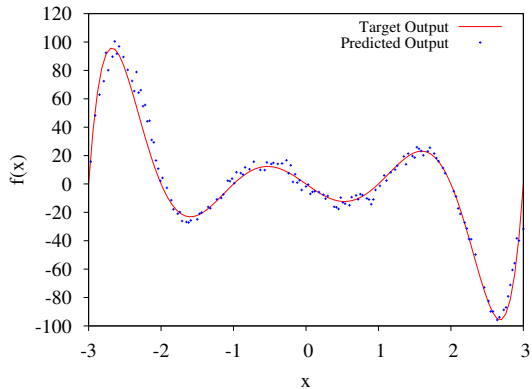}
    \caption{Accuracy of Task-I, using the time series of a laboratory realization of a pendulum as a reservoir, demonstrating the ability of a simple experimental system to execute computational tasks. Here the \textit{frequency encoding} scheme is used.\\
    }
    \label{Fig_expt}
\end{figure}

\bigskip

\noindent
{\bf Comparison of performance with multi-node reservoirs:} \  We have also compared the efficiency of our single-node reservoir with the multi-node reservoirs utilized in earlier studies. As a representative example, we have considered the reservoir of environmentally coupled Lorenz oscillator network \cite{mandal2021achieving} to show the comparison in terms of the similar tasks performed by the two reservoirs. The tasks considered are the attractor reconstruction of chaotic R\"{o}ssler and chaotic Chua systems, and the filtering of a Mackey-Glass time series. 
Table \ref{Tab_comp} lists the accuracy obtained for these tasks performed by both the reservoirs when trained with same training data. These results suggest that our single-node reservoir has the potential to perform better than a reservoir comprised of a large network of dynamical systems.

\begin{table}
    \centering
    \begin{tabular}{|c|c|c|c|}
    \hline
      Reservoir  & \ Task - I \ & \ Task - II \ & \ Task - III \ \\ \hline
 Multi-node Network    &   $10^{-9}$    &   $10^{-6}$    &    $10^{-4}$    \\ \hline
Single  Pendulum   &   $10^{-14}$    &    $10^{-6}$   &     $10^{-11}$   \\ \hline
    \end{tabular}
    \caption{Comparison of accuracy obtained from Reservoir Computing implemented by a multi-node network reservoir and a single-node pendulum reservoir, for three tasks: (I) attractor reconstruction of a chaotic R\"{o}ssler system, (II) attractor reconstruction of a chaotic Chua system, and (III) filtering of a Mackey-Glass time series.}
    \label{Tab_comp}
\end{table}

\medskip

\noindent
{\bf Duffing Oscillator:} \ In order to explore the generality of our results we have also investigated another low-dimensional nonlinear system that can be readily implemented in the laboratory,  the Duffing oscillator. 
Detailed results demonstrating the successful implementation of reservoir computing with a single Duffing oscillator in different dynamical regimes, ranging  from periodic and quasi-periodic to chaotic,
can be found in appendix \textit{A} of the supplementary material. Further these results offer us a test-bed for gauging the comparative performance of systems with different dynamical complexity serving as a single-node reservoir. 
The crucial feature we exploit here is that  transient periodic and quasi-periodic behaviour offers a rich repertoire of temporal sequences, while not suffering from the extreme sensitivity to initial conditions that comes alongside the complexity of chaos. 
So we find that the combination of stability and complexity offered by periodic and quasi-periodic transient dynamics makes this class of dynamical behaviour most suited as a reservoir. 

\medskip

\noindent
{\bf Conclusions:} \ In summary, we have successfully demonstrated that a {\em single simple dynamical system}, such as a pendulum, can be used effectively as a reservoir in Reservoir Computing. Specifically, we exploited the rich dynamics of a driven pendulum for a single-node reservoir to perform complex artificial intelligence tasks. To the best of our knowledge, a single simple pendulum working as an efficient and powerful reservoir is being reported for the first time. 

In this study, we have undertaken two classes of tasks, one processing temporal signals, and the other non-temporal inputs. 
One of the new directions our work suggests is the use of the transient dynamics of non-chaotic nonlinear systems as the ``reservoir'' in single-node reservoir computing, as it offers both stability and complexity. The temporal patterns embedded in the transient dynamics of a nonlinear system can thus provide a rich set of transformations for the {\em readout} layer. We present results from numerical simulations, with the parameters of the dynamical system utilized as reservoir chosen in such a way that it can be easily realized in laboratory experiments.
 
Importantly, this work can also be extended to deal with noisy real-world data sets containing impurities. Further, physical implementations of the idea can be potentially extended to much smaller, faster, and power-efficient systems, for instance, dynamical systems realized with integrated circuit chips. So these ideas can lead to the foundation of powerful machine-learning enabled chips.


\par In conclusion then, we have demonstrated that a single low-dimensional nonlinear dynamical system has remarkable potential for information processing, and can serve as a ``reservoir'' for Reservoir Computing. These results also open up the possibility of other simple dynamical systems for single-node Reservoir Computing, and a wide variety of natural systems can be considered as potential candidates for the reservoir. Thus this work provides a significant step forward towards the broad goal of exploiting intrinsic dynamics of natural systems for intelligent computation.

\bibliography{citations}

\newpage
\newpage
%
\begin{center}
\Huge{\bf Appendix}
\end{center}

\section{Duffing Oscillator}

In order to explore the generality of our results we have investigated another low-dimensional nonlinear system, that can be readily implemented in the laboratory. We have also chosen this system as it offers us a test-bed for gauging the comparative performance of systems with different dynamical complexity. Specifically we implement Reservoir Computing with a single Duffing oscillator serving as a ``reservoir''. The dynamics is given by the evolution equations:

\begin{align}
\dot{x} &= y\\ \nonumber
\dot{y} &= - \delta y - \beta x - \alpha x^3 + f cos(\omega t)
\end{align}

with parameter set chosen as follows:  $\delta = 0.2$, $\beta = -1.0$, $\alpha = 1.0$, $\omega = 1.0$.

    \begin{figure*}
    \centering
    \includegraphics[scale=0.85]{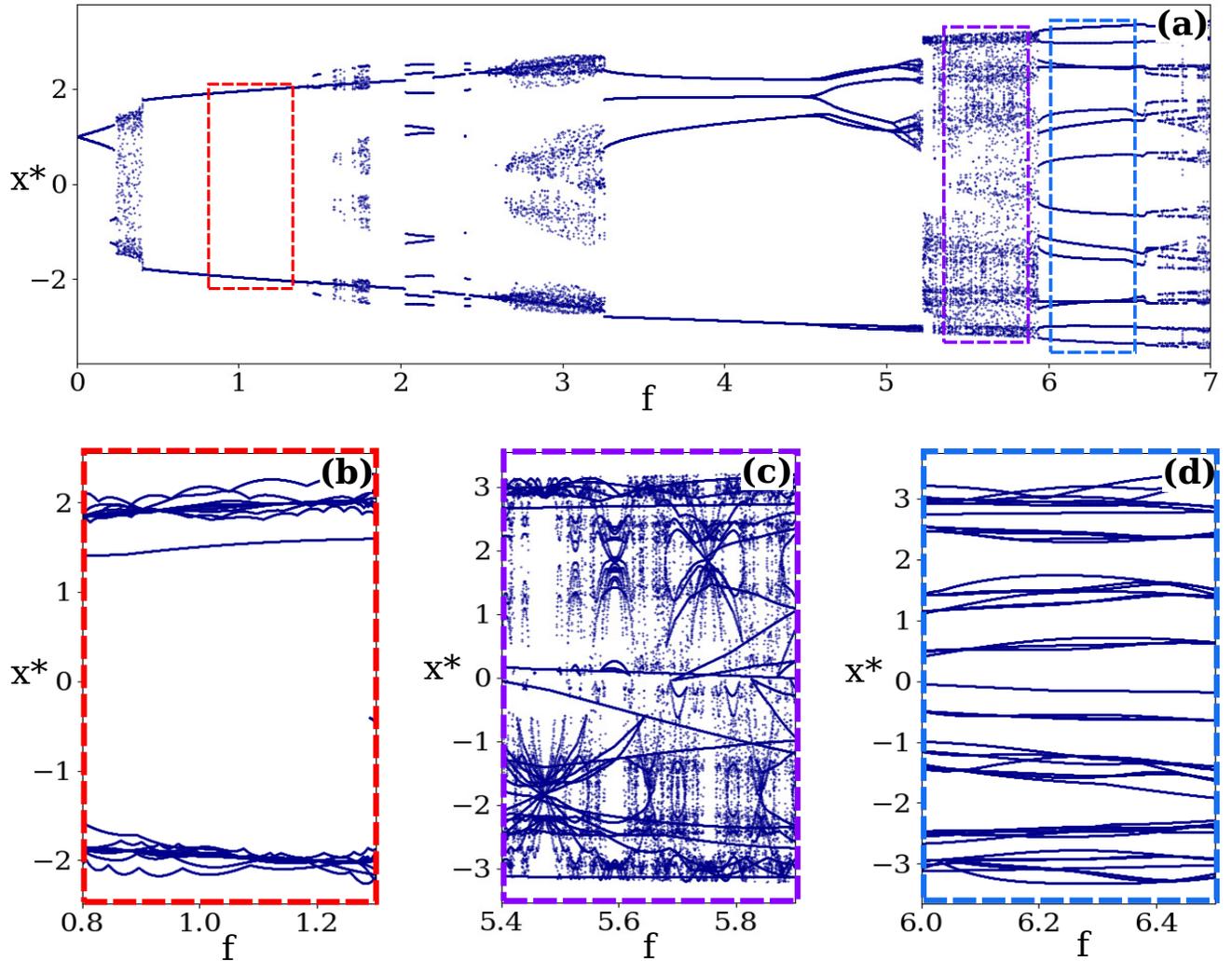}
    \caption{\textbf{(a)} Bifurcation diagram of the asymptotic dynamics of the Duffing oscillator. Panels in the bottom row \textbf{(b - d)} display the bifurcation diagram for the transient counterpart in the periodic, chaotic and quasi-periodic regions respectively, as marked in \textbf{(a)} by dashed rectangles.}
    \label{Fig_bif_duff}
    \end{figure*}

\begin{table*}[h]
\centering

\begin{tabular}{|c|cc|cc|cc|}
\hline
\multirow{2}{*}{} & \multicolumn{2}{c|}{Periodic}     & \multicolumn{2}{c|}{Quasi-periodic}     & \multicolumn{2}{c|}{Chaotic}     \\ \cline{2-7} 
                  & \multicolumn{1}{c|}{Task-I} & Task-II & \multicolumn{1}{c|}{Task-I} & Task-II & \multicolumn{1}{c|}{Task-I} & Task-II \\ \hline
Asymptotic                 & \multicolumn{1}{c|}{$2\times10^{-7}$} & $3\times10^{-3}$ & \multicolumn{1}{c|}{$1\times10^{-7}$} & $1\times10^{-2}$ & \multicolumn{1}{c|}{$1\times10^{0}$} & $2\times10^{0}$ \\ \hline
Transient                 & \multicolumn{1}{c|}{$6\times10^{-12}$} & $2\times10^{-3}$ & \multicolumn{1}{c|}{$2\times10^{-8}$} & $1\times10^{-3}$  & \multicolumn{1}{c|}{$1\times10^{-8}$} & $7\times10^{-3}$ \\ \hline
\end{tabular}


    \caption{Comparison of performance quantified by Root Mean Square Error (RMSE), obtained by Reservoir Computing implemented using a single periodic, quasi-periodic and chaotic Duffing oscillator as a reservoir. The first row presents results obtained using asymptotic dynamics, and the second row presents results obtained using transient dynamics.\\}
    \label{Tab_duff}

\end{table*}

Table \ref{Tab_duff} shows the comparison of the accuracy obtained for two classes of tasks (see main text for details), with the non-temporal task labelled as Task-I and the temporal task labelled as Task-II. Single oscillators with different levels of dynamical complexity are used as the reservoir, ranging from periodic and quasi-periodic, to chaotic. Further we also display the results from Reservoir Computing implemented utilizing asymptotic dynamics shown in the first row, and the results using transient dynamics shown in the second row. 
It is clear that a periodic or quasi-periodic dynamical system serves as a better reservoir in this one-node Reservoir Computing framework, than a chaotic system, where the information loss due to the intrinsic chaos leads to uncertainty in the encoding of inputs. Importantly, it is clearly evident that transient dynamics serves as a better reservoir in Reservoir Computing, with {\em all} systems, ranging from periodic to chaotic performing significantly better when the transient dynamics is utilized. The markedly superior potential of transient dynamics, in comparison to asymptotic dynamics, in tackling computing tasks has not been exploited in earlier work, and presents a new direction for enhanced performance of machine learning based on dynamical systems. \\

\section{Experiment}

\subsection{Components}
\label{appendix:A}

The components used for this experiment are listed and described below.
\begin{enumerate}
    \item A hollow aluminium rod of length $50~cm$ and cross-sectional diameter of $1~cm$, one end attached to a rigid platform to hang from by a pivot. A bob is attached to the other end.
    \item A cylindrical bob of length $6~cm$ and cross-sectional diameter $4.5~cm$, holds two opposite facing propellers aligned in the plane of oscillation. This also contains the control unit of propellers inside with some added weight. The total weight of the bob is $0.5~Kg$ approximately.
    \item Two $A2212/13T~(1000KV)$ brush-less DC motor is attached to the bob with two $10~inch$ (1045) propellers each.
    \item Two $30A$ electronic speed controller (ESC), kept inside the bob, is used for controlling the speed of two motors by a micro-controller.
    \item A micro-controller ($arduino~nano$) attached with bob, is used to receive wireless data and passing it to ESC. This is actually the part of the circuit responsible for generating the driving force function.
    \item $HC~05$ Bluetooth module enables the possibility to receive control input wirelessly without effecting the natural dynamics.
    \item $MPU-6050~GY-521$ gyro sensor is attached to the pendulum near the pivot to collect the angular deflection data of the pendulum with time. This sensor is interfaced with another micro-controller via narrow flexible wires.
    \item $Arduino~Uno$ is used to received and decode the gyro sensor data and to communicate it with the computer for calculation.
    \item An external DC power supply is used to power the whole setup.
\end{enumerate}

\begin{figure*}
    \centering
\begin{tabular}{cc}
\includegraphics[scale=0.605]{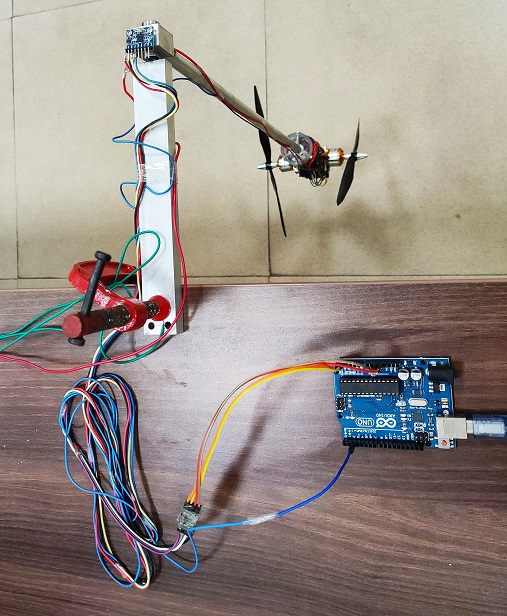} & \includegraphics[scale=0.6]{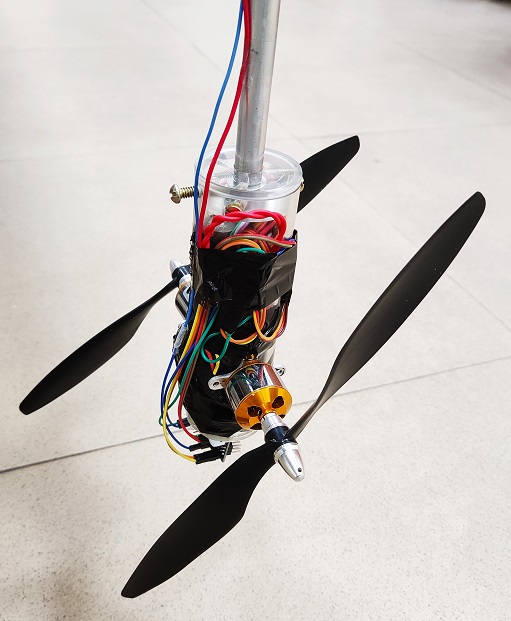}
\end{tabular}
    \caption{Experimental Setup}
    \label{expt_setup}
\end{figure*}

\subsection{Experimental steps}
\label{appendix:B}

\paragraph{Step - I:} We start by setting up the system
with all the components and required circuit connections.

 \paragraph{Step - II:} First, we need to calibrate two BLDC motors. The controllers (ESCs) takes a high frequency square wave signal input and according to its pulse width the speed is decided. So, by the micro-controller we need to generate pulse width modulated (PWM) signal specifying the width in range [0,180]. Basically, $0$ corresponds to no rotation and $180$ to full speed. Now, The speed decides the amount of force exerted ($F$) by the propellers. We need to find out a relation between $f=F/m$ and the input pulse width for PWM.
 
 This is done using a simple setup. Say, for pulse width $p$ a propeller generates a force $F$ and due to that the pendulum rests at an angular displacement $\theta$. In this case, $F = mg~sin(\theta)$ or $f = g~sin(\theta)$. Hence, for any $p$ we can find the value of $f$. Repeating this process with both the motors sufficient number of times with the setup, and fitting the data to straight lines, we can find calibration curves for any value of $f$ for the two motors.
 
 \paragraph{Step - III:} We need to program the micro-controllers according to the requirement of operation. Arduino nano, attached to the bob should receive wireless signal of the ESC inputs and the $\omega$ value to generate the driving force function. Similarly, Arduino Uno, interfacing the gyro sensor should be programmed to sample data at rate defined by $\tau$.
 
 \paragraph{Step - IV:} Running the setup with required inputs and storing the gyro sensor data from Arduino Uno. Reservoir states are generated for both training and testing data inputs.
 
 \paragraph{Step - V:} Training reservoir states are used for regression with their corresponding output.
 
 \paragraph{Step - VI:} Using the optimal output weight evaluated by regression, test reservoir states are used to find the output, and the predicted output is compared with target output.

 \subsection{Drop in performance with the  \textit{amplitude encoding} scheme:}
 Fig.~4 of the letter shows the results from this experiment using the \textit{frequency encoding} scheme. However, there is a significant effect of noise on the performance when the \textit{amplitude encoding} scheme is used. The factors that affect the performance under the \textit{amplitude encoding} scheme are rationalized as follows. To use \textit{amplitude encoding} scheme one need to multiplex the input with the amplitude of force. In the experimental setup, we can control only the pulse width of the ESC input, and two transformations need to be implemented. First, the pulse width information is converted into speed and secondly, according to varying speeds different magnitudes of the reaction force generated by the propellers is exerted on the system. So there are many potential factors affecting the control of input force, such as the electronic or thermal noise effecting the ESCs and air density, ambient temperature, environment's aerodynamics with many other. So there is no direct control over the forcing amplitude, i.e. the value of $f$. On the other hand, the \textit{frequency encoding} scheme simply encodes input information using the frequency of the force $\omega$, on which there is a direct control. Since the frequency can be controlled with a precision of $\sim (\mu s)^{-1}$ by the micro-controller, the frequency-encoding scheme yields better results.

\end{document}